# Substrate-dependent electrical transport in individual single-walled carbon nanotubes grown across $SiO_2$ and hexagonal boron nitride

*Yuanjia Liu* *[a], *Taiki Inoue* *[a], *Yoshihiro Kobayashi* *[a]

a Department of Applied Physics, The University of Osaka,

Suita, Osaka 565-0871, Japan

*Email: liu.y@ap.eng.osaka-u.ac.jp; inoue.taiki@ap.eng.osaka-u.ac.jp;

kobayashi@ap.eng.osaka-u.ac.jp

## Abstract

The electronic transport properties of carbon nanotubes (CNTs) are strongly affected by their surrounding environment, making the underlying substrate a critical factor for device performance. Here, we demonstrate enhanced carrier transport of individual single-walled CNTs on hexagonal boron nitride (hBN) by directly comparing CNT channels on $SiO_2$ and hBN within the same nanotube. This within-tube comparison removes tube-to-tube variability in chirality, diameter, and defect density, allowing the intrinsic substrate effect to be evaluated more reliably. The CNTs were synthesized using gas flow–directed growth, which yields long, well-aligned CNTs without transfer processes, allowing a single nanotube to extend across different substrate regions. Multichannel field-effect transistors fabricated along an individual CNT exhibit clear ambipolar characteristics. CNT channels on hBN consistently exhibit higher field-effect mobility than those on $SiO_2$. In contrast, temperature-dependent transport near the charge neutrality point exhibits thermally activated behavior with similar activation energies (15–20 meV) on both substrates, indicating that the intrinsic small bandgap of CNTs is largely unaffected by the substrate. These results provide direct evidence that hBN enhances low-field carrier transport in CNTs and establish a foundation for the fabrication of high-performance electronics based on hBN-supported CNTs.

Carbon nanotubes (CNTs) [1,2] have attracted sustained attention because of their exceptional electrical, mechanical, and thermal properties [3–5]. Their one-dimensional (1D) structure leads to quantum confinement and reduced scattering, enabling ballistic transport over micrometer scales [6–8]. However, the electronic performance of CNT-based devices is strongly affected by the surrounding environment, including the underlying substrate, surface adsorbates, and fabrication residues [9–11]. Among these, the substrate plays a particularly critical role: it governs charge screening, introduces remote phonon scattering, and determines the level of electrostatic disorder of carriers [12]. Thermally grown $SiO_2$ on Si [13], a typical substrate for nanoelectronics, introduces surface roughness, charge traps, and strong coupling of interfacial phonons, reducing carrier mobility and shortening the electron mean free path [14–17]. In contrast, hexagonal boron nitride (hBN) is an insulating two-dimensional (2D) material with an atomically flat and chemically inert surface [18], low impurity density, and weak phonon coupling [19]. As a result, hBN has been widely used as a substrate for devices based on 2D channel materials, such as graphene and transition-metal dichalcogenides, where van der Waals (vdW) heterostructures improve electrical transport [20–23]. However, whether hBN can directly enhance low-field carrier transport in 1D CNT devices remains less clearly established. Previous study clarified the role of substrate surface polar phonon scattering in CNTs by comparing temperature-dependent resistivity on $SiO_2$ and AlN, but hBN substrates were not examined [24]. Another study demonstrated improved carrier transport in graphene on hBN, but substrate effects in 2D graphene cannot be directly transferred to 1D CNTs [20]. Several studies have fabricated CNTs on hBN as 1D–2D mixed-dimensional vdW heterostructures [25–27] and shown their practical advantages, such as enhanced current-carrying capacity owing to better heat dissipation [28], more stable quantum dot operation [29], reduced hysteresis [30], and improved optical and spectroscopic properties [31,32]. Nevertheless, these studies did not establish an enhancement of low-field carrier mobility for CNTs on hBN substrates. For example, coaxial CNT/hBN heterostructures reduced interface trap density, whereas transferred CNT devices on hBN used as references showed degraded subthreshold swing compared with $SiO_2$, likely because of transfer-induced contamination [30]. A major limitation of these studies is that devices were fabricated from separate CNTs grown on different substrates, making it difficult to distinguish intrinsic substrate effects from tube-to-tube variations in chirality, diameter, or defect density. In addition, variations in fabrication and transfer processes can introduce impurities, strain, or contact resistance differences, which further obscure the intrinsic effects of the substrate. To demonstrate whether hBN enhances carrier transport in CNTs, a direct comparison within the same CNT under nearly identical structural and fabrication conditions is highly desirable.

Herein, we address this limitation by directly comparing CNT-based devices fabricated on $SiO_2$ and exfoliated hBN regions within the same nanotube. To enable this comparison, we employed gas flow–directed growth [33–36], a chemical vapor deposition (CVD) variant in which laminar flow and buoyancy effects guide CNT growth along a uniform direction. This approach enables the direct synthesis of long, aligned CNTs without postgrowth transfer, preserving their intrinsic transport properties and allowing continuous CNTs to extend across different substrate regions on a single chip [35,37,24,38]. Such a configuration is ideal for comparative studies because multichannel field-effect transistors (FETs) can be fabricated along the same CNTs, eliminating tube-to-tube variability. Nanodiamond particles were used as nonmetallic seeds [39,40], enabling the growth of isolated CNTs with minimal contamination, suitable for the fabrication of multichannel devices. Through systematic measurements of temperature-dependent transport, we determined field-effect mobility and activation energies associated with thermally activated conduction. Our results reveal a clear enhancement of field-effect mobility on hBN compared with $SiO_2$, indicating that hBN is a promising underlying substrate for high-performance 1D transport.

We directly synthesized CNTs spanning $SiO_2$ and exfoliated hBN substrate regions using gas flow–directed CVD and fabricated multichannel FETs using standard lithography for direct transport comparison (Fig. 1), with detailed fabrication procedures provided in the Supporting Information. The $SiO_2$ layer on Si substrates had a thickness of 300 nm. The overall device geometry is schematically shown in Fig. 1(a), where a single CNT spans the $SiO_2$ and hBN regions and is contacted by multiple electrodes using a common Si back gate. Optical microscopy further confirmed successful electrode fabrication on both substrates (Fig. 1(b)). The structural and interfacial quality of the CNTs was examined using a combination of microscopy and spectroscopy. Scanning electron microscopy (SEM) images before electrode fabrication reveal that CNTs grew continuously across the $SiO_2$–hBN interface with straight alignment (Fig. 1(c)). Note that SEM imaging was performed under low accelerating voltage and short exposure, followed by postfabrication annealing to recover possible beam-induced damage. [41]. As transport was compared within the same identically processed CNT on hBN and $SiO_2$, the substrate comparison remains valid. The atomic force microscopy (AFM) image in Fig. 1(d) shows that CNTs on hBN exhibit apparent heights of ~1.3 nm, consistent with the typical diameter range of single-walled CNTs, indicating clean and intimate contact with an atomically flat surface. The thickness of the exfoliated hBN flakes is approximately 70–80 nm, as determined using AFM (Fig. S2). Raman spectroscopy of a reference CNT grown separately on the $SiO_2$ substrate and not part of the devices studied herein is provided in Fig. S1. Direct Raman measurements on the device channels were not performed to prevent possible laser-induced heating or contamination during high-power exposure. The spectrum shows a pronounced G-band and negligible D-band contributions, with G/D intensity ratios of 40–50, confirming the high structural quality of CNTs produced via gas flow–directed growth [40]. Raman spectroscopy was used to investigate the hBN substrate. A pronounced and narrow $E_{2g}$ mode centered at ~1370 cm$^{-1}$ is observed, indicating good crystalline quality [42] (Fig. S3). Electrical characterization was conducted under both ambient and high-vacuum conditions using a precision source-measure unit integrated with variable-temperature probe stations. Measurements were performed from 110 to 300 K, enabling the systematic extraction of field-effect mobility and activation energy.

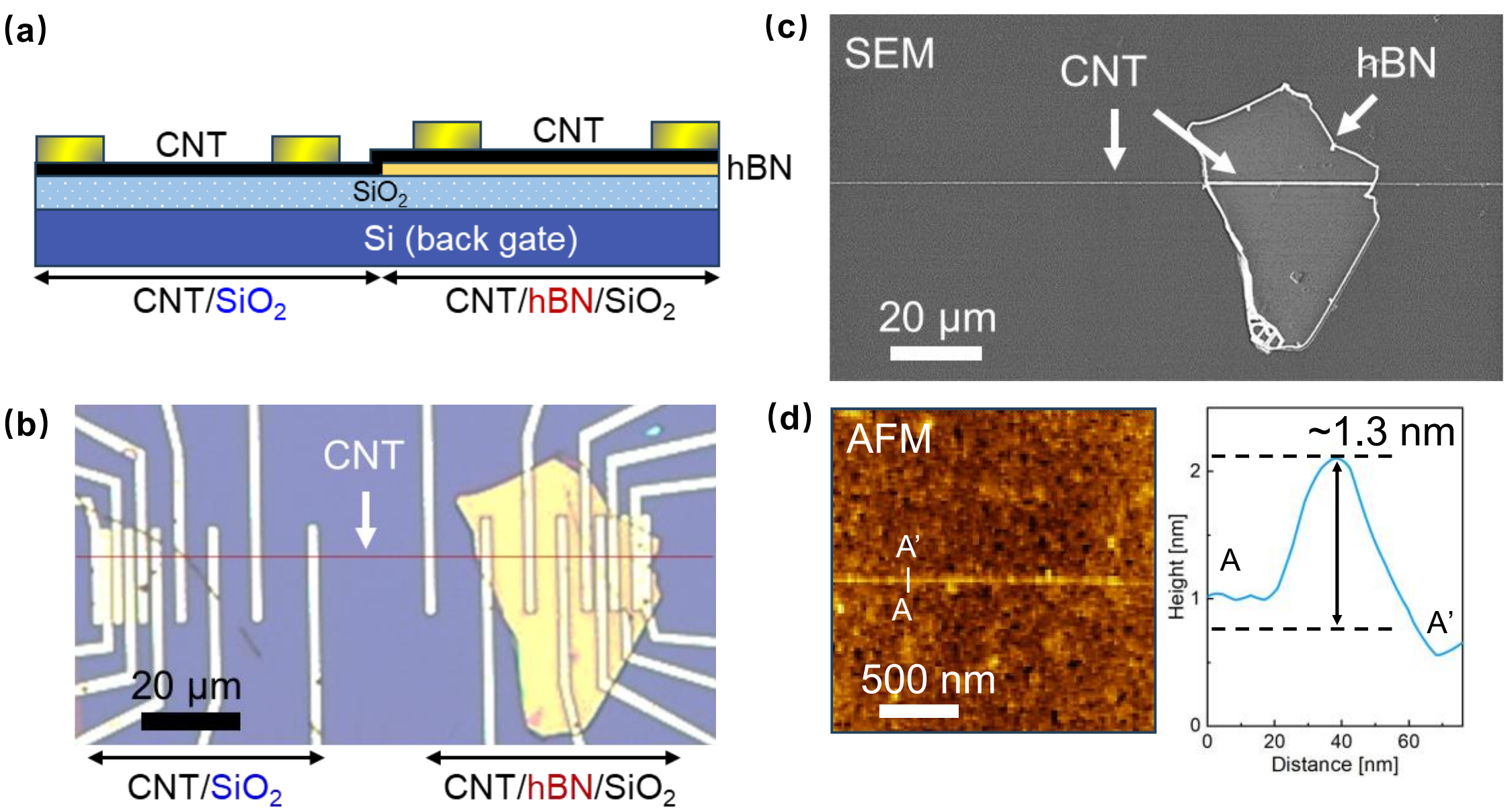

**Figure 1 (a) Schematic of the cross-substrate device geometry, in which multiple channels are defined along a single CNT across $SiO_2$ and hBN regions of the substrate using a common Si back gate. (b) Optical micrograph of the fabricated multichannel CNT devices, showing electrodes patterned on both $SiO_2$ (left) and exfoliated hBN (right) regions. (c) SEM image of the as-grown CNT used for device fabrication. The CNT is well-aligned along the gas-flow direction and spans**

**the hBN region. (d) AFM image of the CNT over hBN, with an apparent height of ~1.3 nm.**

We measured the gate-dependent characteristics of a single CNT spanning both $SiO_2$ and hBN regions at 300 K, which exhibits typical ambipolar behavior [43], as shown in Fig. 2. Although CNT FETs exhibit ambipolar characteristics, the charge neutrality point, corresponding to the minimum conductance, where electron and hole currents are balanced, appears at gate voltages more negative than zero. This ambipolar behavior was attributed to the metal–CNT band alignment at the contacts. The Ti adhesion layer has a relatively low work function, which reduces the injection barrier for electrons. For small-bandgap CNTs, this enables the efficient injection of electrons and holes, leading to symmetric ambipolar transport [44]. In contrast, the negative shift of the charge neutrality point is likely caused by weak n-type charge transfer. Residual amine-containing species from the alkaline developer (tetramethylammonium hydroxide) used during lithography may act as electron donors. Such amine groups are known to donate electrons to CNTs, slightly raising the Fermi level and shifting the transfer characteristics toward negative gate voltages [45]. Beyond these n-type effects, substrate-induced doping also plays a notable role. On $SiO_2$, surface traps and adsorbed species, such as $H_2O$ and $O_2$, are known to introduce p-type doping, whereas hBN provides an atomically flat and chemically inert interface that suppresses such p-type contributions [46]. As a result, in a CNT on hBN, the p-type doping contribution from the substrate is reduced, whereas the contact- and residue-induced n-type doping remains largely unchanged. In multiple devices, the neutrality point is consistently observed at more negative gate voltages than for a CNT on hBN, even though the interfacial environment was cleaner. Because the comparison was performed along the same CNT with identical contacts, this shift was attributed primarily to substrate-related doping effects. As shown in Fig. 2(a), the on-state current on hBN is substantially larger, and the subthreshold slope is steeper, suggesting enhanced electrostatic control and improved contact injection at room temperature. These results directly indicate that hBN is a more favorable supporting substrate for charge transport than conventional $SiO_2$, which introduces charge inhomogeneities and traps that degrade FET performance.

Figures 2(b) and (c) show the temperature-dependent transfer curves of FETs with a 7-μm channel length fabricated on $SiO_2$ and hBN, respectively. Because the present CNT FETs exhibit ambipolar transport, the conventional threshold voltage is not uniquely defined. We therefore analyzed the charge neutrality point, corresponding to the minimum-conductance region, as a threshold-like electrostatic parameter commonly used for ambipolar CNT devices [43]. The charge neutrality point shifts from -2.6 to -6.3 V for CNTs on $SiO_2$ as the temperature increases from 110 to 300 K, whereas CNTs on hBN show a smaller overall variation and become nearly saturated above 170 K (Fig. S4). This difference suggests that thermally activated substrate traps and the resulting charge fluctuations are more pronounced on $SiO_2$, consistent with oxide-interface trap effects [12]. The weaker shift on hBN indicates a more stable interfacial environment with fewer trap-induced charge fluctuations. Figure 2(d) summarizes the field-effect mobility ($\mu_{FE}$) as a function of temperature for the FETs on both substrates. The mobility was extracted from the linear regime using $\mu_{FE} = (L/C_g V_{ds})(dI_d/dV_g)$ [47], where $C_g$ is the gate capacitance per unit length calculated using a cylindrical model. For a single dielectric of thickness $t$ and permittivity $\varepsilon = \varepsilon_0 \kappa$, and for a CNT with diameter $d$, $C_g = 2\pi\varepsilon/\ln(4h/d)$, which corresponds to the classical wire–plane capacitance model [48]. Here, $\varepsilon_0$ is the vacuum permittivity, and $\kappa$ is the relative dielectric constant. For the CNT/hBN channel, a series dielectric model was used to account for the stacked hBN/$SiO_2$ configuration. The effective dielectric constant was evaluated using $\kappa_{\text{eff}} = (t_{\text{hBN}} + t_{\text{SiO}_2})/(t_{\text{hBN}}/\kappa_{\text{hBN}} + t_{\text{SiO}_2}/\kappa_{\text{SiO}_2})$ [49], where $t_{\text{hBN}}$ and $t_{\text{SiO}_2}$ are the thicknesses, and $\kappa_{\text{hBN}}$ and $\kappa_{\text{SiO}_2}$ are the dielectric constants of hBN and $SiO_2$, respectively. The temperature dependence of $\mu_{FE}$ follows a power law, $\mu_{FE} \propto T^{-\gamma}$ [11]. Fitting the data from 110 to 300 K yields $\gamma \approx 1.3$ for $SiO_2$ and $\gamma \approx 1.1$ for hBN. This trend is consistent with phonon-limited transport commonly observed in semiconducting CNTs, where carrier mobility decreases with increasing temperature because of enhanced phonon scattering [11]. The slightly smaller $\gamma$ value on hBN suggests that remote surface phonon scattering is partially suppressed by the atomically flat and nonpolar dielectric surface [50]. Remote phonon scattering arises from the coupling between charge carriers in the CNT and polar optical phonons in the underlying substrate. As $SiO_2$ is a polar oxide with strong surface optical phonon modes, this interaction can limit carrier mobility. In contrast, the weakly polar vdW surface of hBN inhibits this coupling. Therefore, the higher mobility and weaker temperature dependence on hBN are attributed to the reduced contribution of both remote phonon scattering and trap-related scattering. To further elucidate the transport mechanism and assess device reproducibility, additional measurements were performed on CNT channels with lengths of 3 and 5 μm (Fig. S5). The temperature-dependent mobility trends in these devices closely follow those observed for 7-μm channels. A similar substrate-dependent shift in the charge neutrality point is also observed in these devices. The consistent scaling of conductance with channel length indicates diffusive transport dominated by phonon and interfacial scattering [24], from which the mean free path was estimated as 0.4–0.7 μm. This agreement across multiple channel lengths highlights the intrinsic substrate effect rather than device-to-device variations. Moreover, the multichannel device configuration enables the systematic comparison of transport properties within a single CNT, confirming both the reproducibility and physical validity of the observed mobility enhancement on hBN. Nevertheless, the overall mobility values are lower than those reported for optimized semiconducting CNT devices under ideal diffusive transport conditions [47,51]. Such differences were attributed to practical factors, including contact resistance and process-induced disorder.

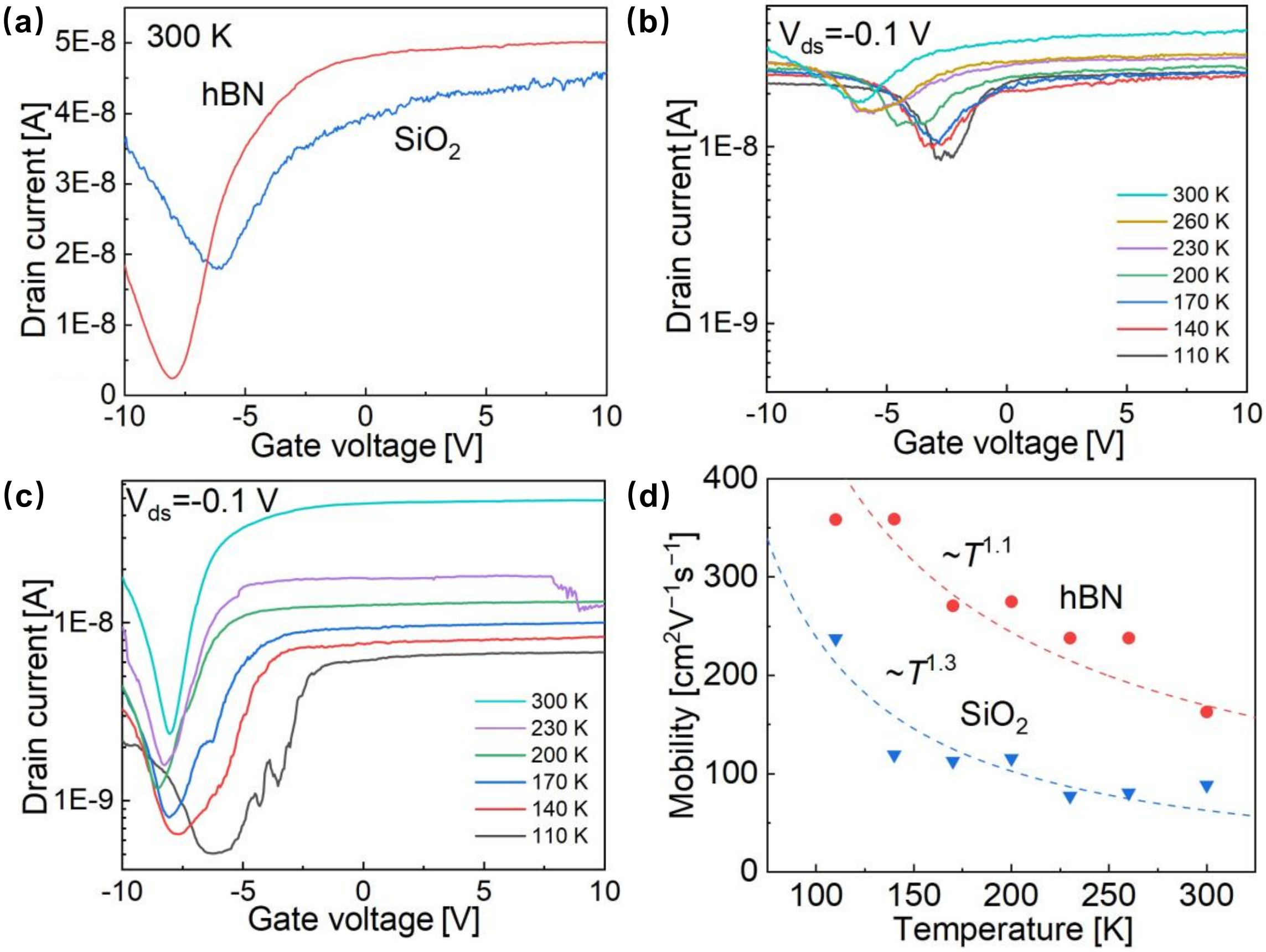

**Figure 2 (a) Transfer characteristics of CNT FETs with a 7-μm channel length on $SiO_2$ and hBN measured at 300 K and $V_{ds} = 0.1$ V. (b,c) Transfer curves for a 7-μm channel on (b) $SiO_2$ and (c) hBN, measured over a temperature range of 110–300 K. (d) Field-effect mobility as a function of temperature for 7-μm CNT channels on $SiO_2$ and hBN. The mobility can be approximated via power-law fitting, shown by the dashed-dotted lines.**

To clarify the role of interfacial morphology in carrier transport, the surface roughness of the two substrates was determined using AFM (Fig. 3). The root-mean-square roughness ($R_q$) was calculated from the height distribution using $R_q = \sqrt{\frac{1}{N}\sum_{i=1}^{N}(z_i - \bar{z})^2}$, where $z_i$ is the measured height at each point and $\bar{z}$ is the mean height. The extracted $R_q$ values were 1.62 nm for $SiO_2$ and 0.89 nm for hBN. Although the hBN surface is substantially smoother than the $SiO_2$ surface, the measured value for hBN remains higher than the subnanometer roughness typically reported for mechanically exfoliated hBN crystals (~0.3 nm) [20]. This discrepancy likely originates from the introduction of contaminants during fabrication, such as amorphous carbon derived from CVD and lithography residues. Despite these effects, the lower roughness of hBN still indicates a flatter and more uniform interface, which suppresses interfacial scattering and charge inhomogeneity at the CNT–substrate interface. This smoother surface is considered one of the key factors contributing to the enhanced carrier mobility on hBN compared with that on $SiO_2$.

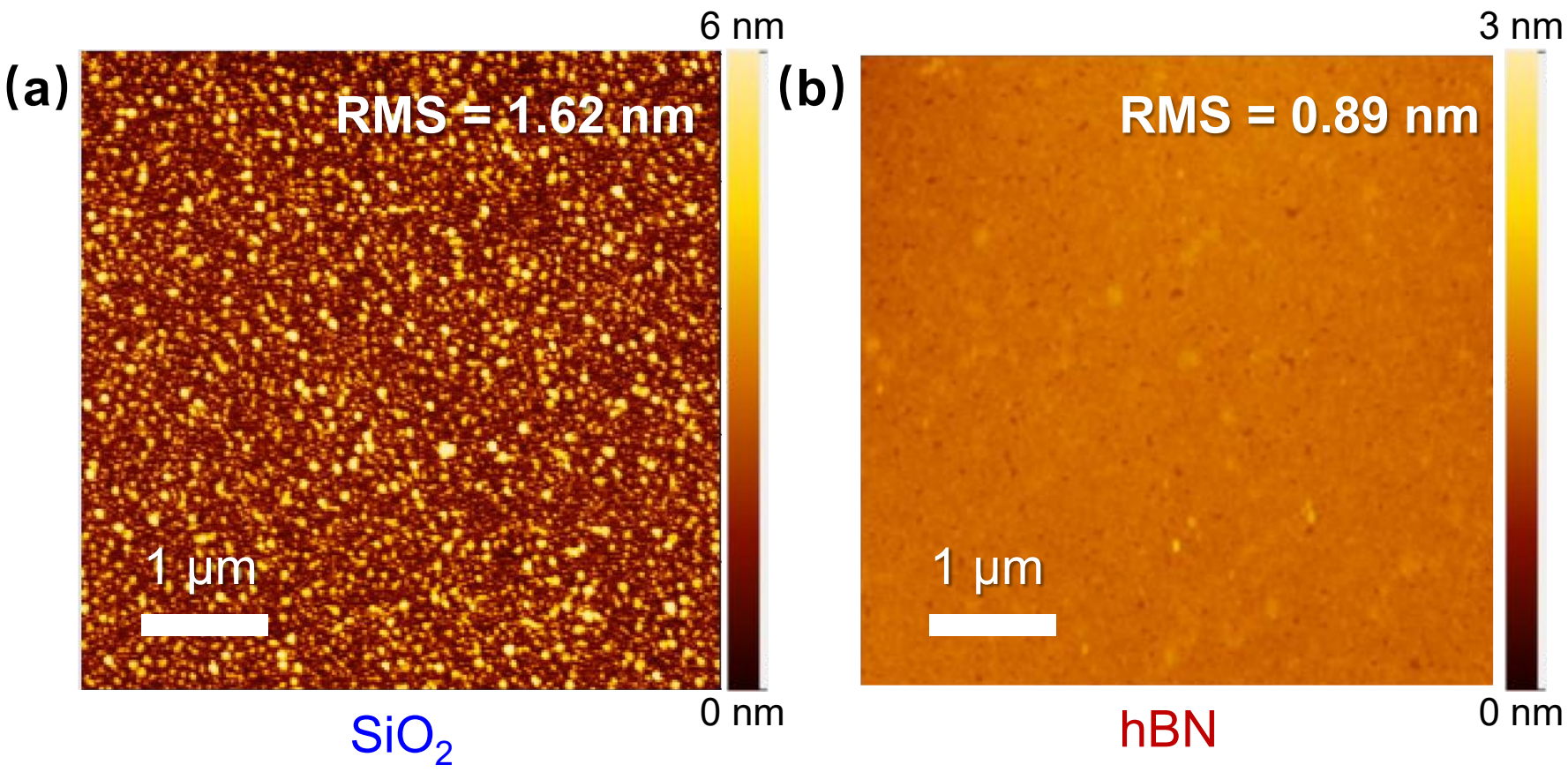

**Figure 3** AFM images of the substrate surfaces showing surface roughness. (a) $SiO_2$ with a root-mean-square (RMS) roughness of 1.62 nm. (b) hBN with a RMS roughness of 0.89 nm.

To further clarify the transport mechanism, we analyzed the temperature dependence of the resistance at the conductance minimum [52] for the same CNT channels on $SiO_2$ and hBN. To eliminate the effects of substrate-dependent charge neutrality shifts, the transfer characteristics were first aligned with respect to the charge neutrality point, and the resistance was then extracted at the minimum-conductance point (Fig. S6). This procedure allowed evaluating the resistance under the same electrostatic conditions, where the Fermi level lies near the middle of the CNT bandgap, thereby allowing the temperature dependence reflects thermally activated transport across the bandgap [52]. Based on the temperature-dependent transfer characteristics shown in Fig. 2(b) and (c), the resistance at the current minimum was extracted and plotted as a function of inverse temperature (Fig. 4(a) and (b)). For both substrates, the data follow an exponential dependence, $R \propto \exp(E_a/k_BT)$, at high temperatures (110–300 K), confirming that the current minimum is indeed governed by thermal activation [52]. The extracted activation energies are approximately 21 meV for the CNT on $SiO_2$ and 16 meV for the CNT on hBN. According to the curvature-induced gap theory, such CNTs exhibit bandgaps scaling as $E_g \propto 1/d^2$, typically on the order of 10–60 meV for diameters around 1 nm [53]. Importantly, although the activation energy on hBN is slightly smaller than that on $SiO_2$, the difference is modest compared to the overall magnitude of the bandgap and does not indicate a substantial modification of the intrinsic CNT band structure, consistent with theoretical calculations [54]. Instead, the similarity of activation energies on the two substrates indicates that the dielectric environment has little effect on the bandgap itself in the present devices [52]. This observation contrasts with the pronounced substrate dependence observed in field-effect mobility and resistivity and clearly distinguishes the roles of intrinsic electronic structure and extrinsic transport processes. At higher temperatures, a deviation from the ideal Arrhenius behavior is observed for the CNT channel on hBN, where the resistance decreases more rapidly than predicted by a purely thermally activated model. This deviation may reflect the thermal population of higher 1D subbands contributing to transport at elevated temperatures, as discussed in prior studies on temperature-dependent CNT conduction [55]. Such behavior is consistent with a cleaner and more homogeneous dielectric interface, which reduces interfacial potential fluctuations and enhances carrier injection efficiency. In contrast, CNT channels on $SiO_2$ are well described by thermally activated transport over the entire temperature range, reflecting stronger interfacial disorder and more pronounced carrier localization near the charge neutrality point.

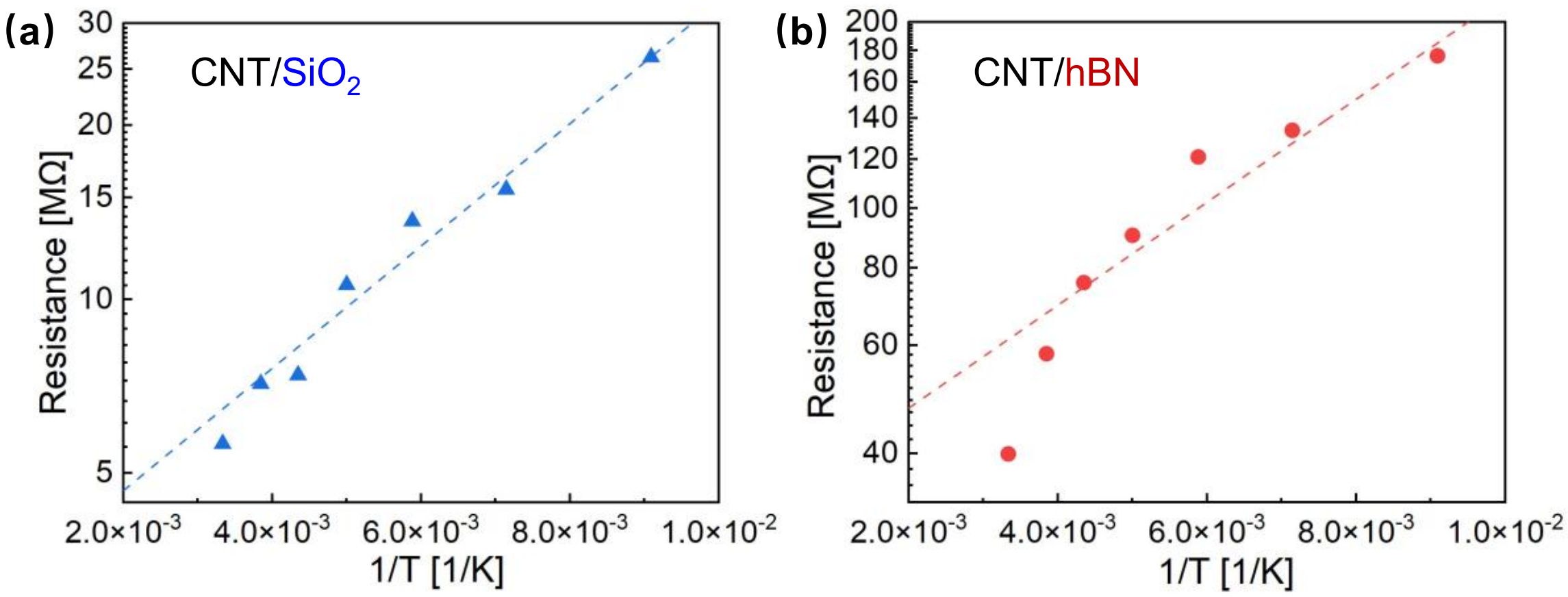

**Figure 4 (a) Plot of resistance for CNT/$SiO_2$, fitted to $R \propto \exp(E_a/k_BT)$, yielding an activation energy of $E_a \approx 21$ meV. (b) Corresponding analysis for CNT/hBN, giving $E_a \approx 16$ meV. The agreement with the thermal activation model confirms that the CNT is a small-bandgap tube.**

These results highlight that the improved transport on hBN originates primarily from extrinsic factors, such as reduced surface roughness, lower trap density, and suppressed remote phonon scattering, rather than from any modification of the intrinsic CNT band structure. They also explain why prior studies observed little improvement on hBN, as transferred CNTs often contained polymer residues and suffered structural damage.

In summary, we demonstrate that hBN enhances low-field carrier transport in individual single-walled CNTs. Gas flow–directed growth enabled long and continuous CNTs to extend across $SiO_2$ and hBN regions without postgrowth transfer, and multichannel FETs fabricated along the same CNT allowed direct substrate comparison under nearly identical structural and contact conditions. Field-effect mobility extracted from the linear regime showed higher values for CNTs on hBN across all measured temperatures, and additional measurements using multiple channel lengths confirmed the same trend, highlighting the reproducibility of this behavior. Temperature-dependent transfer curves exhibited thermally activated behavior near the charge neutrality point. Overall, the present study proves that CNTs supported on hBN exhibit consistently higher field-effect mobility than those on $SiO_2$, while maintaining similar activation energies, indicating that the transport enhancement primarily arises from reduced extrinsic scattering rather than the modification of the intrinsic band structure. The atomically flat and weakly interacting dielectric environment provided by hBN leads to a cleaner electrostatic landscape and improves carrier transport. The present within-tube comparison provides a reliable platform for the quantitative evaluation of substrate effects in CNT-based devices.

## Supplementary material

The supplementary material contains additional figures and supporting information related to device fabrication, electrical measurements, and data analysis. The following files are available free of charge.

APL-Liu yuanjia-SI.pdf

## Acknowledgments

The authors would like to thank Dr. T. Sakata of the Research Center for Ultra-High Voltage Electron Microscopy, The University of Osaka, for assistance in the SEM observation. A part of this work was financially supported by JSPS, KAKENHI (Grant Numbers JP15H05867, JP17H02745, JP24K01297, and JP25K01623), JST (Grant Numbers JPMJCR20B5 and JPMJPR24H1) and Nakatani Foundation. A part of this work was conducted in Advanced Research Infrastructure for Materials and Nanotechnology Open Facilities in The University of Osaka, supported by ARIM Program of the Ministry of Education, Culture, Sports, Science and Technology (MEXT), Japan, Grant Number JPMXP1225OS1006.

## Author declarations

### Conflict of interest

The authors have no conflicts to disclose.

### Ethics approval

This research does not involve human participants or animals.

### Author contributions

Yuanjia Liu: Conceptualization, Methodology, Investigation, Writing-Original Draft preparation;

Taiki Inoue: Conceptualization, Funding Acquisition, Resources, Methodology, Supervision, Writing-Review & Editing;

Yoshihiro Kobayashi: Funding Acquisition, Resources, Methodology, Supervision, Writing-Review & Editing.

## Data availability statement

The data that support the findings of this study are available from the corresponding author upon reasonable request.

Highlight Figure

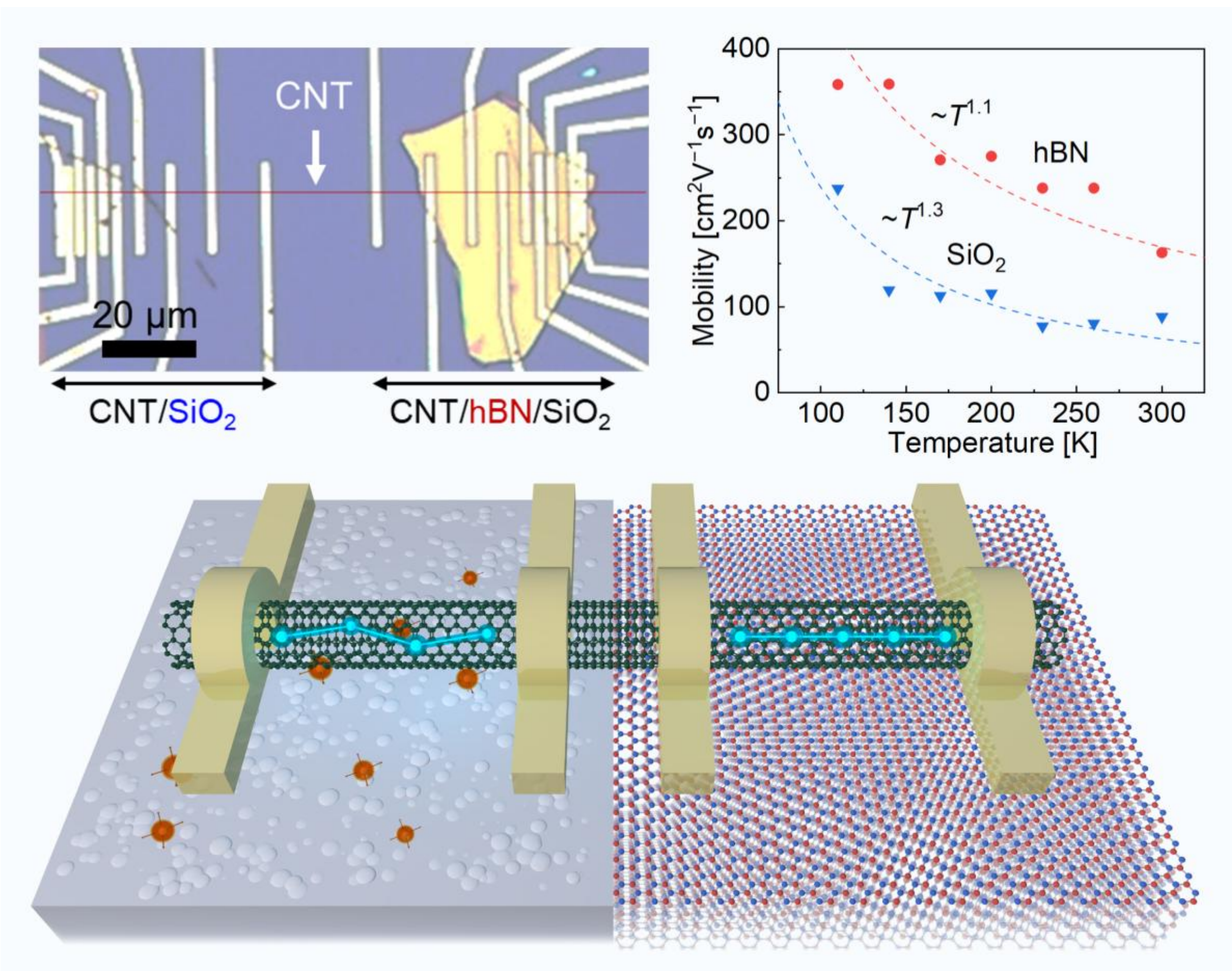

Supporting information

# Substrate-dependent electrical transport in individual single-walled carbon nanotubes grown across $SiO_2$ and hexagonal boron nitride

*Yuanjia Liu* *[a], *Taiki Inoue* *[a], *Yoshihiro Kobayashi* *[a]

a Department of Applied Physics, The University of Osaka,

Suita, Osaka 565-0871, Japan

*Email: liu.y@ap.eng.osaka-u.ac.jp; inoue.taiki@ap.eng.osaka-u.ac.jp;

kobayashi@ap.eng.osaka-u.ac.jp

## S1. Experimental

### 1.1 CNT growth

A ~1 cm × 1 cm $SiO_2$/Si wafer was used as the substrate for carbon nanotube (CNT) growth. Fresh adhesive tape was first used to mechanically exfoliate hexagonal boron nitride (hBN) flakes from bulk hBN crystals (HQ-graphene). The same piece of hBN-bearing tape was then laminated onto one half of the $SiO_2$/Si wafer to transfer hBN flakes and to mask that region from nanodiamond (ND) deposition, which serves as the growth seeds for CNTs. The other half, which remained bare $SiO_2$, was drop-cast with an ethanol dispersion of ND. The ND suspension was allowed to dry in air. The wafer was placed on a hot plate at 90 °C for 2.5 min. The tape was then peeled off to leave a half-hBN, half-ND substrate: one half covered by transferred hBN flakes, the other half seeded with ND on bare $SiO_2$. Growth conditions followed our previous gas flow-directed protocol [37]. Substrates were loaded into a horizontal quartz-tube CVD reactor. The tube was preheated to 600 °C. Samples were pushed to the hot zone and held at 600 °C in air for 10 min to remove tape residues and surface contaminants, as well as to downsize ND seeds appropriate for CNT growth [38]. The furnace was then brought to the synthesis temperature in the range 850–950 °C under a continuous $H_2$ (3%)/Ar flow. CNT growth was initiated by introducing ethanol vapor using $H_2$ (3%)/Ar bubbling through an ethanol tank kept at ~40 °C in combination with a main $H_2$ (3%)/Ar stream. The growth process was conducted at atmospheric pressure, and the partial pressure of ethanol was estimated to be ~1.4 kPa. Growth proceeded for ~30 min. After growth, the gas was switched to Ar and the system was cooled to room temperature. Under these conditions, laminar flow and buoyancy guided CNTs along the gas-flow direction. As a result, long and well-aligned CNTs were obtained, some of which spanned from the ND-seeded $SiO_2$ region into the hBN region, enabling subsequent multi-channel device fabrication on the same CNT for cross-substrate transport studies.

### 1.2 Alignment markers and electrodes fabrication

To facilitate scanning electron microscopy (SEM)-based CNT localization, alignment markers were first patterned via photolithography. Layered resist structures were formed by sequentially spin-coating hexamethyldisiloxane (HMDS), LOR-3A, and OFPR, with each layer baked at the appropriate temperature (120 °C for HMDS, 150 °C for LOR-3A, and 110 °C for OFPR). Exposure was performed using a UV lamp through a photomask, followed by development in a tetramethylammonium hydroxide (TMAH) developer for 60 s. Au alignment markers were then deposited using a quick coater (SC-701MkII ADVANCE, Sanyu Electron), producing conductive markers with sufficient optical and SEM contrast. Lift-off was achieved in N-methyl-2-pyrrolidone (NMP) at 90 °C for 1 h.

After CNT mapping by SEM, electrode patterns were fabricated on CNTs by electron-beam lithography (ELS-BODEN-OU4801, ELIONIX). A diluted ZEP520A resist (1:3 in anisole) was spin-coated at 2000 rpm for 60 s and baked at 180 °C for 3 min. Patterns were written at 150 keV, ensuring sub-100 nm precision. Post-exposure, samples were developed in ZED-N50 for 90 s before drying under nitrogen. Metal electrodes were deposited in an Anelva L-043E electron beam evaporation system, where the chamber base pressure was ~$10^{-4}$ Pa. A Ti adhesion layer (50 nm) was deposited first, followed by Au with a typical thickness of 450–950 nm. Thickness and deposition rates were monitored by quartz crystal microbalance. Lift-off was again performed in NMP at 90 °C for 1 h. After the lift-off process, the samples were annealed at 300 °C for 30 min in a 3 % $H_2$ in Ar to remove resist residues and improve the metal–CNT contact quality.

### 1.3 Characterization

Raman spectroscopy was employed to assess the structural quality of the CNTs. Spectra were acquired using a Raman microscope (RAMANtouch, Nanophoton) with a 532 nm excitation laser. The characteristic G-band (~1590 $cm^{-1}$) and D-band (~1350 $cm^{-1}$) were monitored to evaluate defect density. The G/D intensity ratio served as a key indicator of CNT crystallinity and defect concentration. Atomic force microscopy (AFM; AFM5000, Hitachi High-Tech) was used to determine CNT diameters and to inspect surface morphology, including the thickness of exfoliated hBN flakes and the step height between hBN and $SiO_2$ regions. AFM imaging was performed in tapping mode under ambient conditions. SEM (S-4800, Hitachi High-Tech, accelerating voltage 1 kV, emission current 10 μA, acquisition time <30 s, high vacuum) was utilized to map CNT positions across the substrate, enabling precise localization for subsequent lithography. SEM imaging also confirmed the successful spanning of individual CNTs across $SiO_2$ and hBN regions and was used to verify electrode alignment after fabrication.

Electrical characterization was carried out using an electrical characterization system (Keysight B2912A precision source/measure unit) connected to four complementary probe stations. Under an optical microscope, probes were brought into direct contact with the electrode pads, allowing the assignment of source, drain, and gate electrodes and enabling standard current–voltage (*I*–*V*) characterization. For variable-temperature transport, a vacuum-compatible probe station (Model E-4, Riko-Boeki) was employed. This system incorporates a turbomolecular pumping unit, providing a base pressure of ~$10^{-3}$ Pa, and a liquid nitrogen cooling system, enabling temperature control from 110 K to 300 K.

## S2. Supplementary figures

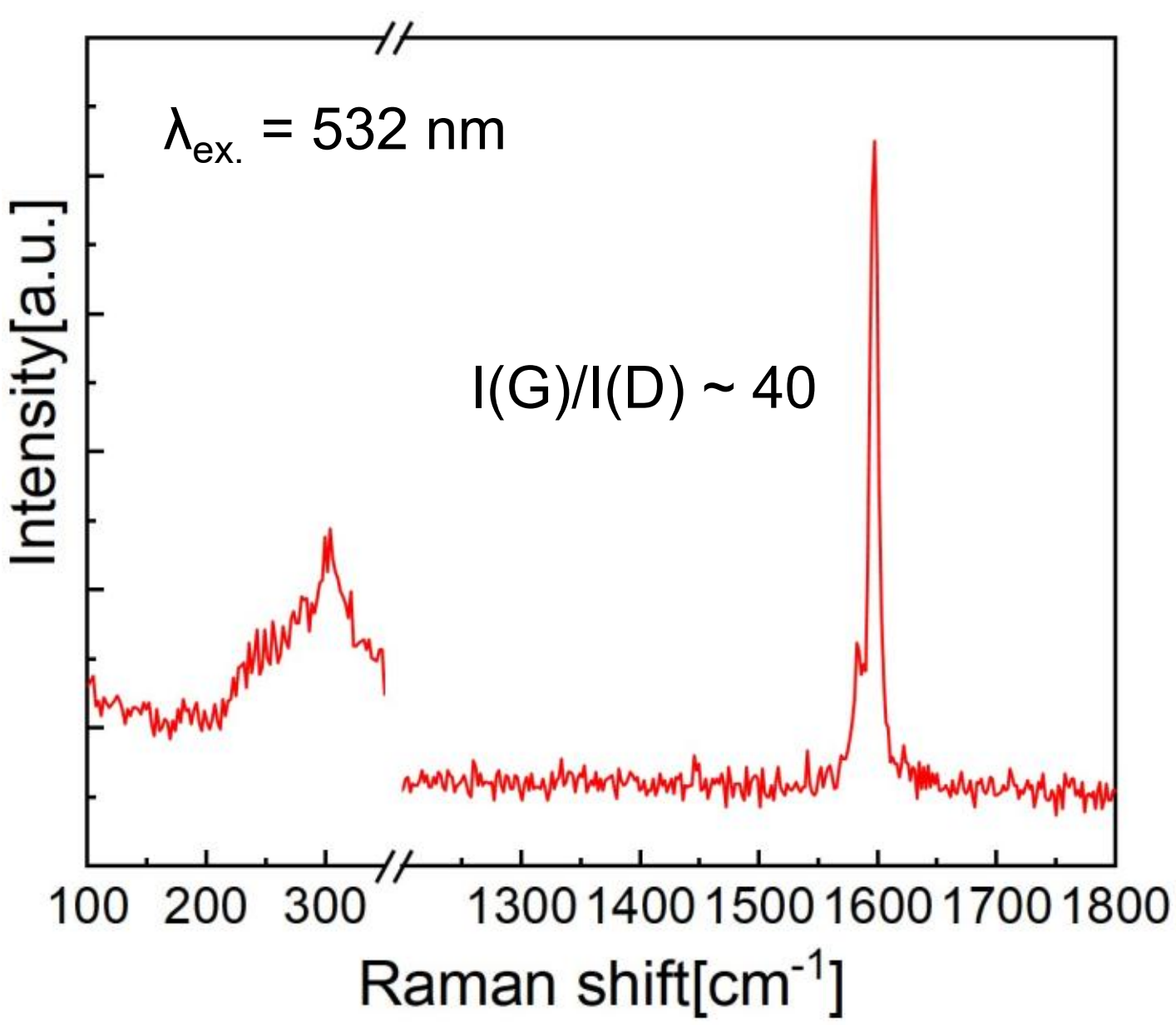

**Figure S1** Raman spectrum of a CNT grown directly on a $SiO_2$ substrate.

**Alt text:** Low- and high-wavenumber regions of Raman spectrum measured with 532 nm excitation. The high-wavenumber spectrum shows a peak near 1590 $cm^{-1}$, and the G-band to D-band intensity ratio is approximately 40.

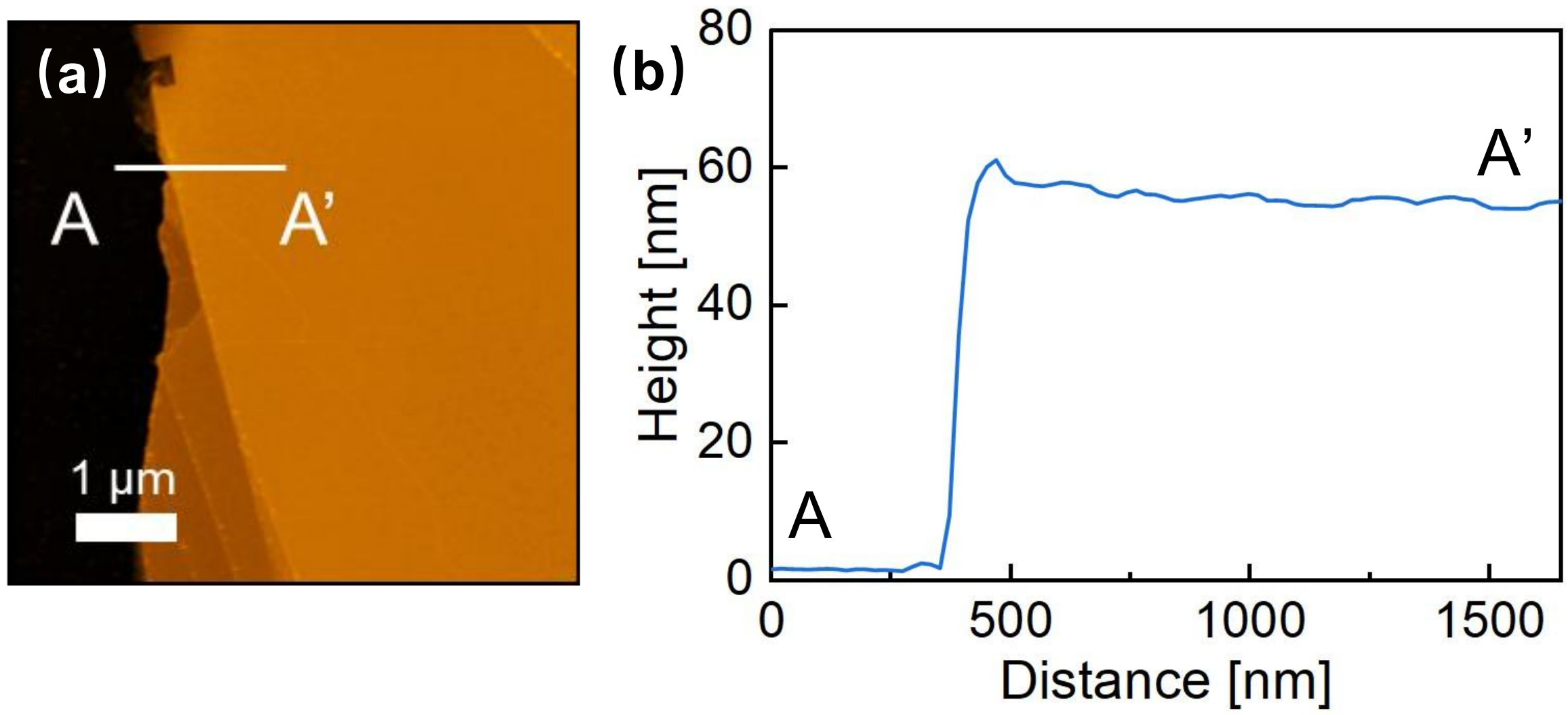

**Figure S2** AFM characterization of an exfoliated hBN flake. (a) Topographic image showing a clear step edge. (b) Corresponding height profile.

**Alt text:** AFM topographic image and corresponding height profile across a boundary between a substrate and an hBN flake. The height profile shows a step height of approximately 60 nm.

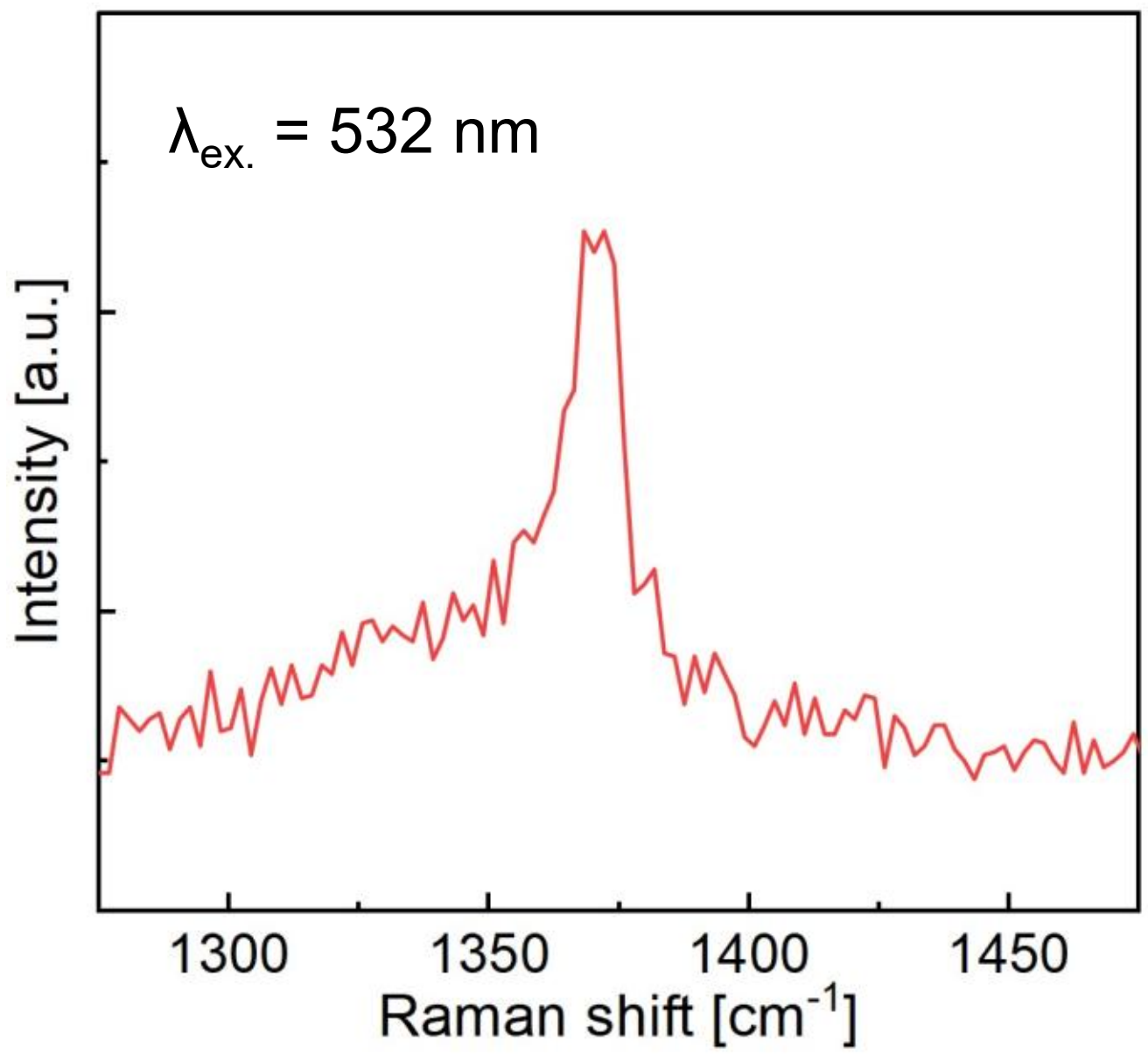

**Figure S3** Raman characterization of the hBN substrate.

**Alt text:** Raman spectrum measured with 532 nm excitation. The spectrum shows a peak near 1370 $cm^{-1}$.

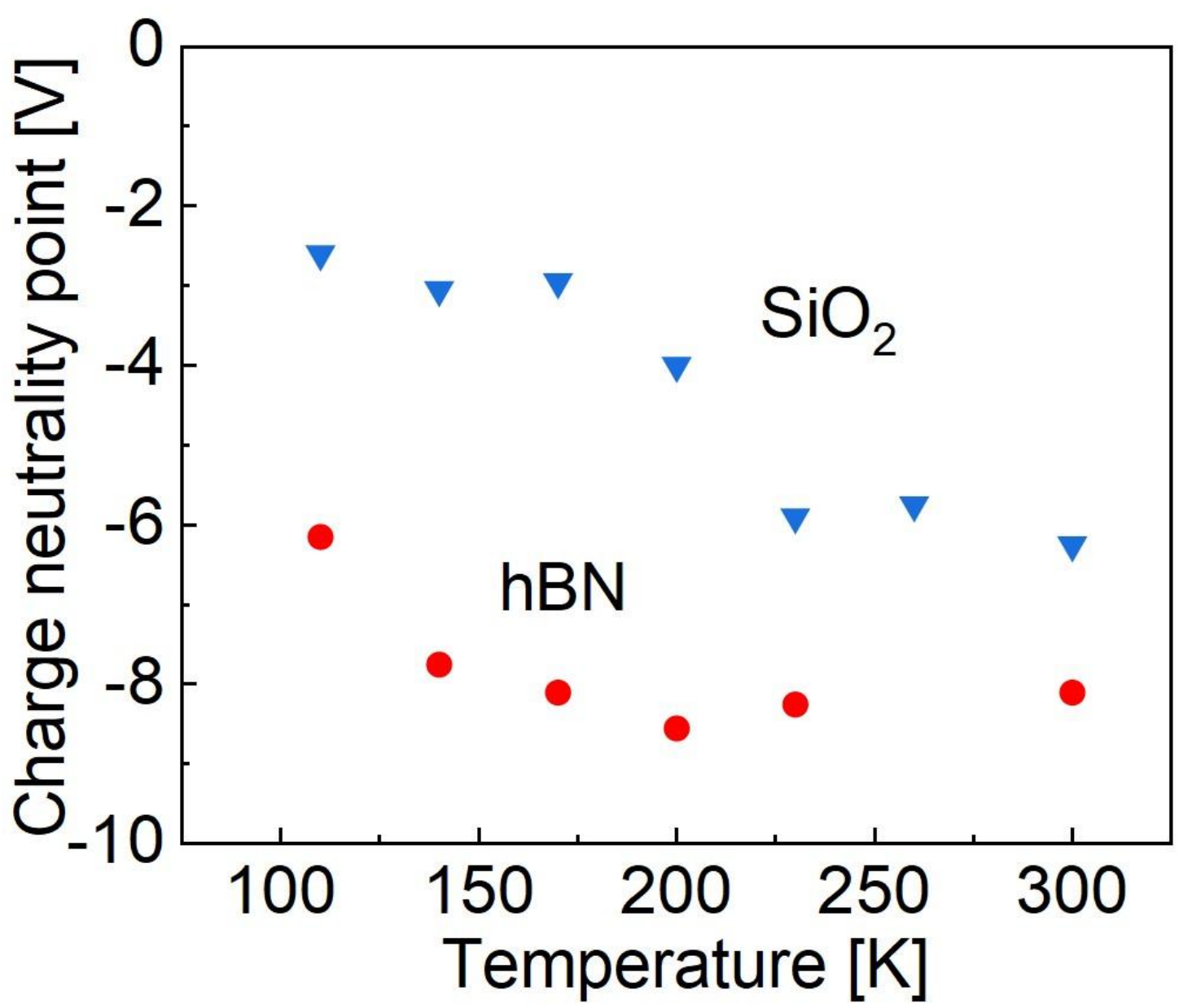

**Figure S4** Temperature dependence of the gate voltage at the charge neutrality point for CNT channels on $SiO_2$ and hBN. The data are extracted from the transfer characteristics shown in Fig. 2(b) and 2(c) for a device with a channel length of 7 μm. The charge neutrality point is defined as the gate voltage corresponding to the minimum conductance in the transfer characteristics at each temperature.

**Alt text:** Plot of charge neutrality point as a function of temperature, comparing data for $SiO_2$ and hBN.

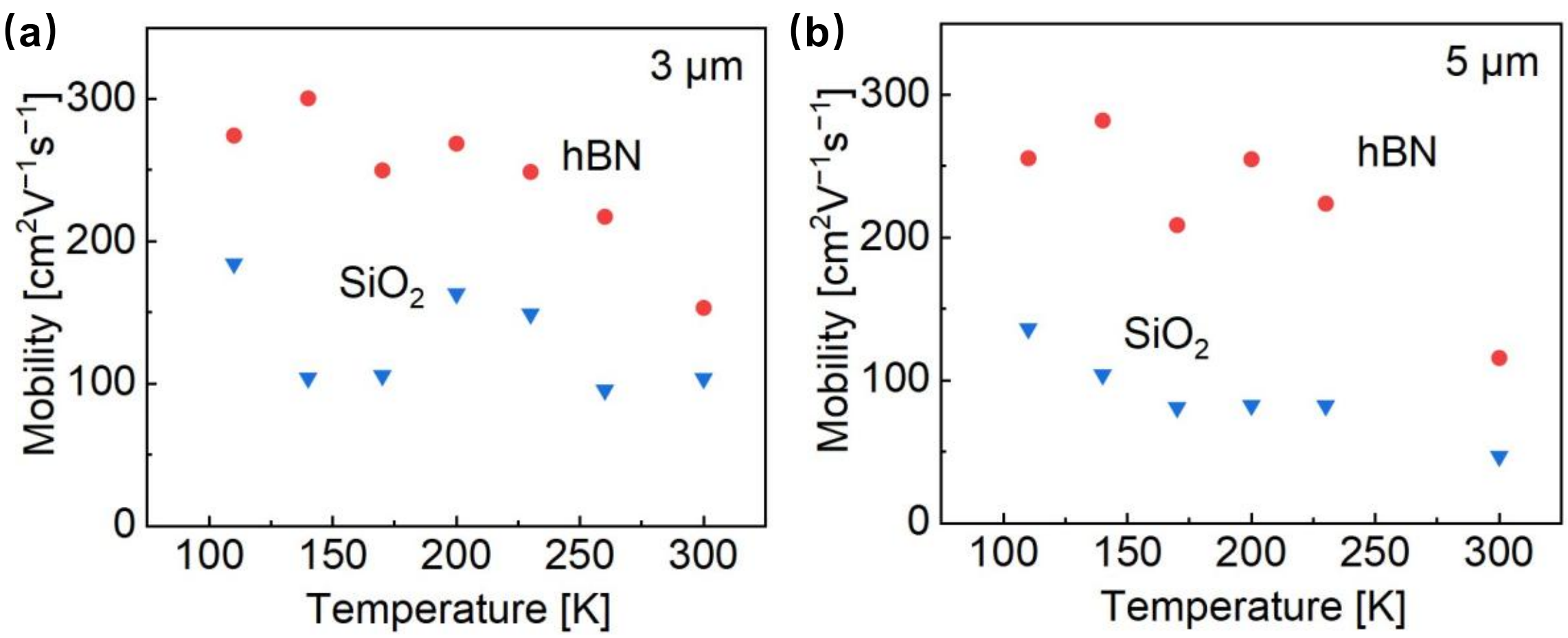

**Figure S5** Temperature-dependent mobility of CNT devices with a 3 μm channel length on $SiO_2$ and hBN substrates. (b) Temperature-dependent mobility of CNT devices with a 5 μm channel length on $SiO_2$ and hBN substrates.

**Alt text:** Plot of mobility as a function of temperature, comparing data for $SiO_2$ and hBN.

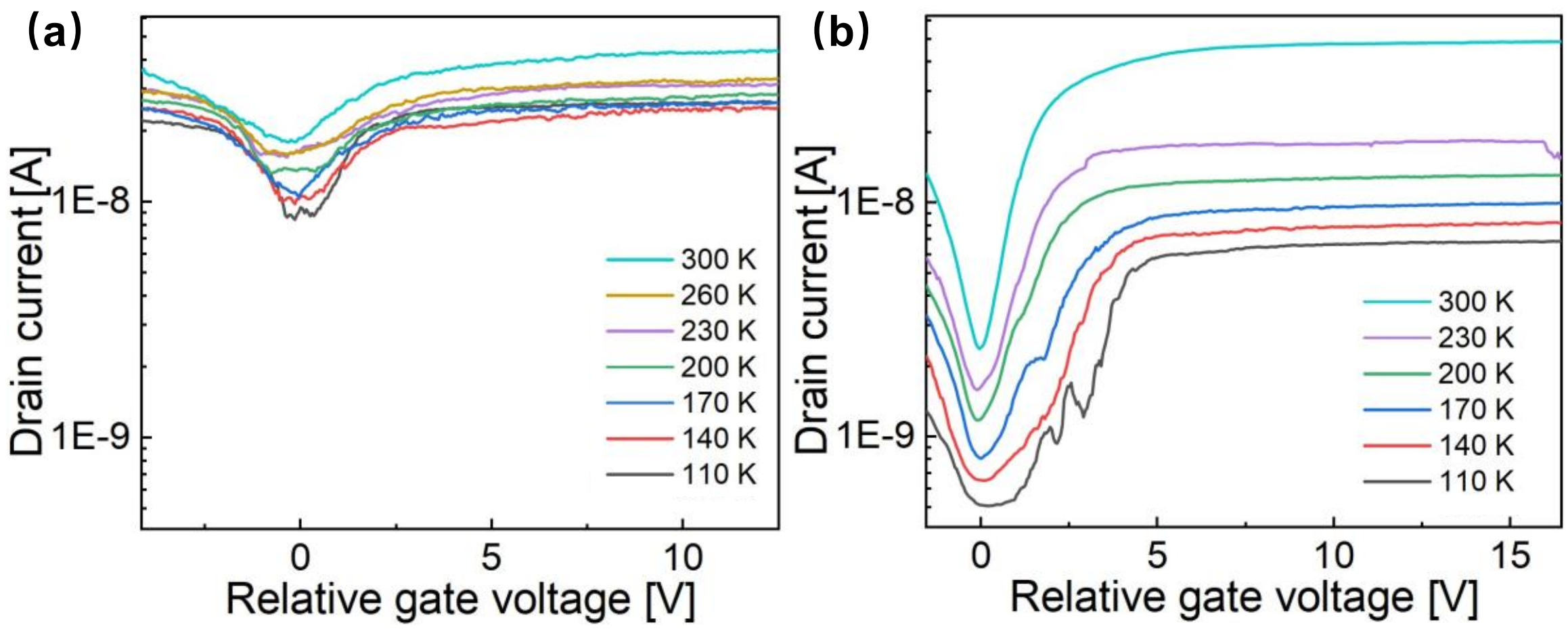

**Figure S6** Temperature-dependent transfer characteristics of CNT FETs with a 7 µm channel length after alignment of the charge neutrality point. To facilitate comparison, the gate-voltage axis for each curve is shifted so that the conductance minimum (charge neutrality point) is located at 0 V for all temperatures. (a) CNT channel on $SiO_2$. (b) CNT channel on hBN. Measurements were performed from 110 K to 300 K.

**Alt text:** Transfer curves plotted as drain current versus relative gate voltage at temperatures from 110 K to 300 K.